\documentstyle[prl,aps,epsfig,floats]{revtex}

\begin{document}

\twocolumn[\hsize\textwidth\columnwidth\hsize\csname
@twocolumnfalse\endcsname

%\preprint{hep-th/0112254, DF/IST-01.2002}
\draft
\tighten

\title{\bf Black hole collision with a scalar particle in 
three dimensional anti-de Sitter spacetime} 
\author{Vitor Cardoso, Jos\'e P. S. Lemos}
\address{CENTRA, Departamento de F\'{\i}sica,
	      Instituto Superior T\'ecnico,\\ 
               Av. Rovisco Pais 1, 1096 Lisboa, Portugal,\\
          E-mails: vcardoso@fisica.ist.utl.pt, lemos@kelvin.ist.utl.pt}
\setlength{\footnotesep}{0.5\footnotesep}
\maketitle

%%%%%%%%%%%%%%%%%%%%%%%%%%%%%%%%%%%%%%%%%%%%%%%%%

\begin{abstract}
\noindent
We study the collision between a BTZ black hole 
and a test particle coupled to a scalar
field. We compute the power spectrum,
the energy radiated and the plunging waveforms for this 
process. We show that for late times the signal is dominated 
by the quasinormal ringing. In terms of the AdS/CFT 
correspondence the bulk gravity process maps into a thermal state, 
an expanding bubble and gauge particles decaying into bosons of
the associated operator. These latter thermalize in a timescale 
predicted by the bulk theory. 

\end{abstract}

\vskip1pc]

%%%%%%%%%%%%%%%%%%%%%%%%%%%%%%%%%%%%%%%%%%%%%%%%%
\noindent
\section{ Introduction}
\vskip 3mm
%%%%%%%%%%%%%%%%%%%%%%%%%%%%%%%%%%%%%%%%%%%%%%%%%

Anti-de Sitter (AdS) spacetime has been considered of fundamental
meaning within high energy elementary particle physics, specially in
supersymmetric theories of gravity such as 11-dimensional supergravity
and M-theory (or string theory).  The dimension $d$ of AdS spacetime
is a parameter which can have values from
two to eleven, and where the other spare
dimensions either are joined as a
compact manifold $\cal M$ into the whole spacetime to yield ${\rm
AdS}_d\times {\cal M}^{11-d}$ or 
receive a Kaluza-Klein treatment.  AdS spacetime appears as the background
for black holes solutions and it also plays a
further crucial role since it is the near-horizon geometry, separated
by a (soft) boundary from an otherwise asymptotically flat spacetime,
of some black solutions \cite{horostrometal}.  In addition, by taking
low energy limits at strong coupling and through group theoretic
analysis, Maldacena conjectured a correspondence between the bulk of
AdS spacetime and a dual conformal field gauge theory (CFT) on the
spacetime boundary itself \cite{maldacenaconjecture}.  A concrete
method to implement this correspondence is to identify the extremum of
the classical string theory action $I$ for the dilaton field $\varphi$,
say, at the boundary of AdS, with the generating functional $W$ of the
Green's correlation functions in the CFT for the operator $\cal O$
that corresponds to ${\varphi}$
\cite{gubserklebanovpolyakov}, 
$
I_{\varphi_0(x^\mu)}\,=
\, W[\varphi_0(x^\mu)]\,, 
$
where $\varphi_0$ is the value of $\varphi$ at the AdS boundary and the
$x^\mu$ label the coordinates of the boundary. The motivation for this
proposal can be seen in the reviews
\cite{aharonyetalklebanovhorowitzreview}.  In its strongest form the
conjecture requires that the spacetime be asymptotically AdS, the
interior could be full of gravitons or containing a black hole.  The
correspondence realizes the holographic principle, since the bulk is
effectively encoded in the boundary, and is also a strong/weak
duality, it can be used to study issues of strong gravity using weak
CFT or CFT issues at strong coupling using classical gravity in the
bulk.

A particular important AdS dimension is three.  In AdS$_3$ Einstein
gravity is simple, the group of isometries is given by two copies of
$SL(2,R)$, it has no propagating degrees of freedom, is
renormalizable, it allows for the analytical computation of many
physical processes extremely difficult or even impossible in higher
dimensions, it belongs to the full string theory compactification
scheme \cite{aharonyetalklebanovhorowitzreview}, the dual CFT$_2$ is
the low-energy field theory of a D1-D5-brane system which can be
thought of as living on a cylinder (the boundary of AdS$_3$)
\cite{maldacenastrom}, and it contains the BTZ black hole.  The BTZ
black hole is of considerable interest, not only because it can yield
exact results, but also because one hopes that the results can
qualitatively be carried through to higher dimensions.  Several
results related to the BTZ black hole itself and to the AdS/CFT
correspondence have been obtained
\cite{keski,danielsson,mat,balasu}.  The AdS/CFT
mapping implies that a black hole in the bulk corresponds to a thermal
state in the gauge theory \cite{wittenbanksetal}. Perturbing the black
hole corresponds to perturbing the thermal state and the decaying of
the perturbation is equivalent to the return to the thermal state.
Particles initially far from the black hole correspond to a blob (a
localized excitation) in the CFT, as the IR/UV duality teaches
\cite{susskindwitten}.  The evolution towards the black hole represents
a growing size of the blob with the blob turning into a bubble
traveling close to the speed of light \cite{danielsson}.

In this  work we extend some of the previous results and we study in
detail the collision between a BTZ black hole and a scalar particle.
Generically, a charged particle falling towards a black hole emits
radiation of the corresponding field. In higher dimensions it also
emits gravitational waves, but since in three dimensions there is no
gravitational propagation in the BTZ case there is no emission. 
%(this also simplifies the problem since gravitational radiation
%involves gauge problems and energy definition, and even in
%higher dimensions the infall of a scalar particle and consequent
%scalar emission should give qualitatively similar results to the
%gravitational wave case).  
Thus, a scalar particle falling into a BTZ
black hole emits scalar waves. This collision process is important
from the points of view of three-dimensional dynamics and 
of the AdS/CFT conjecture. Furthermore, one can
compare this process with previous works, since there are exact
results for the quasinormal mode (QNM) spectrum of scalar 
perturbations which are known to govern their decay 
at intermediate and late times
\cite{HHcardosolemos}.

The phenomenon of radiation emission generated from an infalling
particle in asymptotically flat spacetimes has been studied by several
authors \cite{zerillietal} and most recently in \cite{loustoprice},
where the results are to be compared to full scale numerical
computations for strong gravitational wave emission of astrophysical
events \cite{baker} which will be observed by the GEO600, 
LIGO and VIRGO projects. A scalar infalling particle as a model for
calculating radiation reaction in flat spacetimes has been considered in 
\cite{burko}.
Many of the
techniques  have been developed in connection
to such spacetimes.  Such an
analysis has not been carried to non-asymptotically flat spacetimes,
which could deepen our understanding of these kind of events, and of
Einstein's equations.  In this respect, for the mentioned reasons, 
AdS spacetimes are the most promising candidates.  
As asymptotically flat spacetimes
they provide well defined conserved charges and positive 
energy theorems, which makes them a good
testing ground if one wants to go beyond flatness. However, 
due to different boundary conditions it raises new
problems. The first one is that since the natural boundary conditions are 
boxy-like all of the generated radiation will eventually fall into
the black hole, thus infinity has no special meaning in
this problem, it is as good a place as any other, i.e., one can
calculate the radiation passing at any radius $r$, for
instance near the horizon.  Second, in contrast to asymptotically flat
spacetimes, here one
cannot put a particle at infinity (it needs an infinite amount of
energy) and thus the particle has to start from finite $r$. This has
been posed in \cite{loustoprice} but was not fully solved when applied
to AdS spacetimes.

%%%%%%%%%%%%%%%%%%%%%%%%%%%%%%%%%%%%%%%%%%%%%%%%%
\noindent
\section{ Formulation of the problem and basic equations}
\vskip 3mm
%%%%%%%%%%%%%%%%%%%%%%%%%%%%%%%%%%%%%%%%%%%%%%%%%

We consider a small test particle of mass $m_0$ and charge 
$q_0$, coupled to a massless scalar field $\varphi$, moving along 
a radial timelike
geodesic outside a BTZ black hole of mass $M$. The metric
outside the BTZ black hole is
\begin{equation}
ds^{2}= f(r) dt^{2}- \frac{dr^{2}}{f(r)}-
r^{2}d\theta^{2} \,,
\label{lineelement}
\end{equation}
where, $f(r)=(-M+\frac{r^2}{l^2})$, $l$ is the AdS radius 
(G=1/8; c=1). 
The horizon radius is given by
$r_+=M^{1/2}l$.
We treat the scalar field as a perturbation,
so we shall neglect the back reaction of the field's stress
tensor on the metric (this does not introduce large errors 
\cite{gleiser}).
If we represent the particle's worldline by
$x^{\mu}=x_p^{\mu}(\tau)$, with $\tau$ the proper
time along a geodesic, then the interaction action $\cal I$ is 
\begin{eqnarray}
&{\cal I}
=-\frac{1}{8 \pi} \int g^{1/2} \varphi _{;a} \varphi ^{;a} d^3y-
\nonumber\\
&
m_0 \int (1+q_0 \varphi)(-g_{ab}\dot{x}^a \dot{x}^b)^{\frac{1}{2}} 
d\tau\,, &
\label{action}
\end{eqnarray}
and thus the scalar field satisfies
the inhomogeneous wave equation
$
\Box \varphi= -4 \pi m_0q_0 \int
\delta^3(x^{\mu}-x_p^{\mu}(\tau))(-g)^{-1/2}d\tau\,,
$
where $g$ is the metric determinant and $\Box$ denotes
the covariant wave operator.
As the particle moves on a timelike geodesic, we have
\begin{equation}
\dot{t}_p=\frac{\cal E}{f(r_p)}\;\;,\dot{r}_p=-({\cal E}^2-f(r_p))^{1/2}\,,
\label{geodesic}
\end{equation}
where $\dot{ }\equiv d/d\tau$, 
and $\cal E$ is a conserved energy parameter.
We shall be considering the test particle initially at rest
at a distance $r_0$ (where ${\cal E}^2=f(r_0)$) and at $\theta_p=0$.
Expanding the field as 
\begin{equation}
\varphi(t,r,\theta)=\frac{1}{r^{1/2}}\phi(t,r)\sum_m e^{im\theta} \,,
\label{decomposition}
\end{equation}
where $m$ is the angular momentum 
quantum number, the wave equation is given by
(after an integration in $\theta$)
\begin{eqnarray}
&\frac{\partial^{2} \phi(t,r)}{\partial r_*^{2}} -
\frac{\partial^{2} \phi(t,r)}{\partial t^{2}}-V(r)\phi(t,r)=
\nonumber\\
&
-\frac{2q_0 m_0 f}{r^{1/2}}(\frac{dt}{d\tau})^{-1}\delta(r-r_p) \,,&
\label{waveeq2}
\end{eqnarray}
with
$
V(r)=\frac{3r^2}{4 l^4} - \frac{M}{2 l^2}-\frac{M^2}{4 r^2}+\frac{m^2}{l^2}
 - \frac{Mm^2}{r^2}\,,
%\label{potentialscalar}
$
and 
$
r_*=-M^{-1/2}{\rm arcoth}(r\,M^{-1/2}).
$
%%%%%%%%%%%%%%%%%%%%%%%%%%%%%%%%%%%%%%%%%%%%%%%%%
\noindent
\section{ The initial data and boundary conditions}
\vskip 3mm
%%%%%%%%%%%%%%%%%%%%%%%%%%%%%%%%%%%%%%%%%%%%%%%%%

In the case we study, and in contrast to asymptotically flat 
spacetimes where initial data can be pushed to infinity \cite{zerillietal}, 
initial data must be provided. 
Accordingly, we take the Laplace transform $\Phi(\omega,r)$ of
$\phi(t,r)$ to be 
\begin{equation}
\Phi(\omega,r)=\frac{1}{(2\pi)^{1/2}}\int_{0}^{\infty}
e^{i \omega t}\phi(t,r)dt.
\label{laplace}
\end{equation}   
Then, equation (\ref{waveeq2}) may be written as 
\begin{equation}
\frac{\partial^{2} \Phi(r)}{\partial r_*^{2}} +
\left\lbrack\omega^2-V(r)\right\rbrack\Phi(r)=S
+ \frac{i\omega \phi_0}{(2 \pi)^{1/2}}\,,
\label{waveeq3}
\end{equation}
with,
$
%\begin{equation}
S=-\frac{f}{r^{1/2}}(\frac{2}{\pi})^{1/2}\frac{1}{\dot{r}_p} 
e^{i\omega t} \,
%\label{source}
%\end{equation}
$
being the source function, 
and $\phi_0$ is the initial value of $\phi(t,r)$ satisfying
$
%\begin{eqnarray}
%&
\frac{\partial^{2} \phi_0(r,m)}{\partial r_*^{2}} 
-V(r)\phi_0(r,m)=
%&
%\nonumber \\
%&
-\frac{2f}{r^{1/2}}(\frac{dt}{d\tau})_{r_0}^{-1}\delta(r-r_p) \,.
%&
%\label{initial}
%\end{eqnarray}
$
We have rescaled $r$, $r\rightarrow\frac{r}{l}$, and measure
everything in terms of $l$, i.e., $\phi$, $r_+$ and $\omega$
are to be read, $\frac{1}{l^{1/2}q_0 m_0}\phi$, 
$\frac{r_+}{l}$ and $\omega l$, respectively.
One can numerically solve the equation for the initial
data $\phi_0$ by demanding regularity at both the horizon
and infinity (for a similar problem see \cite{wald}).
In Fig. 1, we show the form of $\phi_0$ for a typical case
$r_+=0.1$,  
$r_0=1$, and for three different values of $m$, 
$m=0,\,1,\,2$. 
Other cases like $r_+=1,\,10,\,...$ and several values of $r_0$ can be
computed.  Large black holes have a direct interpretation in the
AdS/CFT conjecture. The results for large or small black holes are
nevertheless similar, as we have checked.  As a test for the
numerical evaluation of $\phi_0$, we have checked that as $r_0
\rightarrow r_+$, all the multipoles fade away, i.e., $\phi_0
\rightarrow 0$, supporting the no-hair conjecture (that all the
multipoles go to zero).

\vskip 3mm
\centerline{\epsffile{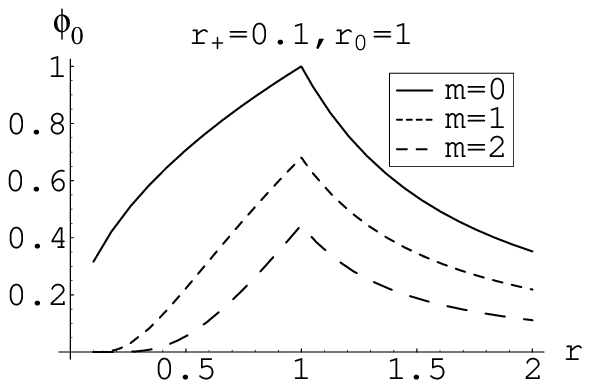}} 
{\noindent {\small Figure 1. 
Initial data $\phi_0$ for a BTZ black hole with $r_+=0.1$, and with the
particle at $r_0=1$, for several values of $m$, the angular quantum
number. }} 
\vskip 3mm

To solve equation (\ref{waveeq3}) one has to impose physically
sensible boundary conditions, appropriate to AdS spacetimes.
In our case the potential diverges at infinity,
where $\phi_0$ vanishes, so we impose reflective boundary conditions
\cite{avis} there, i.e., $\Phi=0$ at infinity.  It has been common
practice to set $\Phi\sim F(\omega) e^{-iwr_*} $ near the horizon,
meaning ingoing waves there. This is an allowed boundary condition as
long as $\phi_0$ vanishes there. However, if $\phi_0$ does not vanish
there, one has to be careful in defining boundary conditions at the horizon.
When dealing with AdS spacetimes this detail is crucial to extract 
the correct information, and it has been overlooked when one deals 
with asymptotically flat spacetimes as in \cite{loustoprice}. 
Equation (\ref{waveeq3})
together with the source term $S$ 
%(\ref{source}) 
allow us to conclude that near the horizon
$
\Phi \sim G(\omega)e^{i\omega r_*}+F(\omega) e^{-i\omega r_*}+
\frac{i \phi_0}{(2\pi)^{1/2}\omega }\,.
$
Since we want waves going down the black hole,  we shall require
\begin{equation}
\Phi \sim F(\omega) e^{-i\omega r_*}+
\frac{i \phi_0}{(2\pi)^{1/2}\omega }\,\,\,, r\rightarrow r_+
\label{boundarybehavior2}
\end{equation}
%%%%%%%%%%%%%%%%%%%%%%%%%%%%%%%%%%%%%%%%%%%%%%%%%
\noindent
\section{ Green's function solution}
\vskip 3mm
%%%%%%%%%%%%%%%%%%%%%%%%%%%%%%%%%%%%%%%%%%%%%%%%%

To proceed we must find a solution to equation (\ref{waveeq3}) through
a Green's function analysis.  A standard treatment \cite{loustoprice}
invokes contour integration to calculate the integrals near the
horizon.  There is no need for this here, by demanding regularized
integrals the correct boundary conditions appear in a natural way (see
\cite{poisson} for a regularization of the Teukolsky equation).  Let
$\Phi^{\infty}$ and $\Phi^{H}$ be two independent solutions of the
homogeneous form of (\ref{waveeq3}), satisfying:
$\Phi^H \sim e^{-i\omega r_*}\,,r \rightarrow r_+ \,$;
$\Phi^H \sim A(\omega) r^{1/2}+B(\omega)r^{-3/2}\,,r \rightarrow \infty \,$; 
$\Phi^{\infty} 
\sim C(\omega)e^{i\omega r_*}+D(\omega)e^{-i\omega r_*}\,,r 
\rightarrow r_+  \,$;
$\Phi^{\infty} \sim 1/r^{3/2}\,,r \rightarrow \infty\,$.
Define $h^H$ through $dh^H/dr_*=-\Phi^H$ and $h^{\infty}$ 
through $dh^{\infty}/dr_*=
-\Phi^{\infty}$.
We can then show that $\Phi$ given by 
\begin{eqnarray}
&
\Phi=\frac{1}{W} \left[ \Phi^{\infty} \int _{-\infty}^{r}\Phi^H S dr_* +
\Phi^H \int _{r}^{\infty}\Phi^{\infty} S dr_* \right]+
\nonumber\\
&\frac{i\omega}{(2\pi)^{1/2} W } \left[ \Phi^{\infty} 
\int _{-\infty}^{r}h^H \frac{d\phi_0}{dr_*} dr_* +\right.
\nonumber \\
&
\Phi^H \int _{r}^{\infty}h^{\infty} \frac{d\phi_0}{dr_*} dr_* + 
\nonumber \\
& 
\left. (h^{\infty}\phi_0 \Phi^H-h^H\phi_0\Phi^{\infty})(r)
\right]\,, 
\label{Phi}
\end{eqnarray}
is a solution to (\ref{waveeq3}) 
and satisfies
the boundary conditions.
The Wronskian $W= 2i\omega C(\omega)$ is a constant.
Near infinity, we get from (\ref{Phi}) that
\begin{eqnarray}
&
\Phi(r\rightarrow\infty)=\frac{1}{W} 
\left[ \Phi^{\infty}_{(r\rightarrow\infty)} \int _{-\infty}^{\infty}\Phi^H S dr_* \right]+
\nonumber\\
&\frac{i\omega}{(2\pi)^{1/2} W } \left[ \Phi^{\infty}_{(r\rightarrow\infty)} 
\int _{-\infty}^{\infty}h^H \frac{d\phi_0}{dr_*} dr_* +\right.
\nonumber \\
& 
\left. (h^{\infty}\phi_0 \Phi^H-h^H\phi_0\Phi^{\infty})(r\rightarrow\infty)
\right]\,.
\label{Phiinf}
\end{eqnarray}
Now, in our case, this is just zero, as it should, because both $\Phi^{\infty},
\phi_0 \rightarrow 0$, as $r\rightarrow \infty$.
However, if one is working with asymptotically flat space, as in \cite{loustoprice},
where $\Phi^{\infty}\rightarrow e^{i\omega r_*}$ at infinity, we get (recalling
that $\phi_0 \rightarrow 0$): 
\begin{eqnarray}
&
\Phi(r\rightarrow\infty)=\frac{1}{W} 
\left[ \Phi^{\infty}_{(r\rightarrow\infty)} \int _{-\infty}^{\infty}\Phi^H S dr_* \right]+
\nonumber\\
&\frac{i\omega}{(2\pi)^{1/2} W }  \Phi^{\infty}_{(r\rightarrow\infty)} 
\int _{-\infty}^{\infty}h^H \frac{d\phi_0}{dr_*} dr_* \,,
\label{Phiflat}
\end{eqnarray}
and where each integral is well defined.
In particular, integrating by parts the second integral can be put 
in the form
\begin{eqnarray}
&
\int _{-\infty}^{\infty}h^H \frac{d\phi_0}{dr_*} dr_* =
\left[h^H \phi_0 \right]_{-\infty}^{\infty}+\int _{-\infty}^{\infty}\Phi^H \phi_0{dr_*}\nonumber\\
&=\frac{i\phi_0 e^{-i\omega r_*} }{\omega}(r\rightarrow -\infty)+\int _{-\infty}^{\infty}\Phi^H \phi_0{dr_*}.
\label{Phiflat2}
\end{eqnarray}
Here, the final sum converges, but not each term in it.
Expression (\ref{Phiflat2}) is just expression (3.15) in \cite{loustoprice}, although
it was obtained imposing incorrect boundary conditions and not well defined regularization
schemes. Due to the fact that the initial data vanishes at infinity,
the results in \cite{loustoprice} are left unchanged.
In this work, we are interested in computing the wavefunction $\Phi(r,\omega)$ 
near the horizon ($r\rightarrow r_+$). In this 
limit we have
\begin{eqnarray}
&
\Phi(r\sim r_+)=\frac{1}{W} \left[ \Phi^H 
\int _{r_+}^{\infty}\Phi^{\infty} S dr_* \right]+&
\nonumber \\
&
\frac{i\omega}{(2\pi)^{1/2}W }
\Phi^H\left[ \int _{r_+}^{\infty}\Phi^{\infty} \phi_0 dr_*-
(h^{\infty}\phi_0)(r_+) \right]+
\nonumber \\
& 
+\frac{i\phi_0(r_+)}{(2\pi)^{1/2}\omega }\,,&
\label{Phi22}
\end{eqnarray}
where an integration by parts has been used.
Fortunately, one can obtain an exact expression for $\Phi^{\infty}$ 
in terms 
of hypergeometric functions \cite{HHcardosolemos}.
The results for $\Phi^{\infty}$ and $W$ are 
\begin{equation}
\Phi^{\infty}=
\frac{1}{r^{3/2}(1-M/r^2)^{\frac{i\omega}{2M^{1/2}}}}
F(a,b,2,\frac{M}{r^2})\,,
\label{solution}
\end{equation}
\begin{equation}
W=2\,i\,\omega\,
\frac{2^{\frac{i\omega}{M^{1/2}}}\Gamma(2)\Gamma(-\frac{i\omega}{M^{1/2}})}
{M^{3/4}\Gamma(1+i \frac{m-\omega}{2\sqrt{M}})
\Gamma(1-i \frac{m+\omega}{2\sqrt{M}})}.
\label{solutionwronskian}
\end{equation}
Here, $a=1+i \frac{m-\omega}{2\sqrt{M}}$ and $b=1-i \frac{m+\omega}{2\sqrt{M}}$.
So, to find $\Phi$ we only have to numerically integrate (\ref{Phi22}).
We have also determined $\Phi^{\infty}$ numerically by 
imposing the boundary conditions above. 
The agreement between the numerical 
computed $\Phi^{\infty}$ and
(\ref{solution}) was excellent. To find $\phi(t,r)$ one must 
apply the inverse Laplace transformation to $\Phi(\omega,r)$. 
Integrating the wave equation in this spacetime is simpler
than in asymptotically flat space, in the sense that, due to the
boundary conditions at infinity the solution is more stable,
and less effort is needed to achieve the same accuracy.
Using a similar method to that in \cite{loustoprice}, we estimate
the error in our results to be limited from above by 0.5\%.
%%%%%%%%%%%%%%%%%%%%%%%%%%%%%%%%%%%%%%%%%%%%%%%%%
\noindent
\section{ Numerical results for the waveforms and spectra}
\vskip 3mm
%%%%%%%%%%%%%%%%%%%%%%%%%%%%%%%%%%%%%%%%%%%%%%%%%

To better understand the numerical results, we first point out that
the QNM frequencies for this geometry, calculated by
Cardoso and Lemos \cite{HHcardosolemos} (see also \cite{birmingham}
for a precise relation between these QNM frequencies and the poles of the 
correlation functions on the CFT side) are 
\begin{equation}
\omega_{\rm QNM}=\pm m -2iM^{1/2}(n+1).
\label{clequation}
\end{equation}
In Fig. 2 we show the waveforms for the $r_+=0.1\,,\,\,r_0=1$ black hole, 
as a function of the advanced null-coordinate $v=t+r_*$. 
This illustrates in a beautiful way 
that QNMs govern the late time behavior of the waveform. 
\vskip 3mm
\centerline{\epsffile{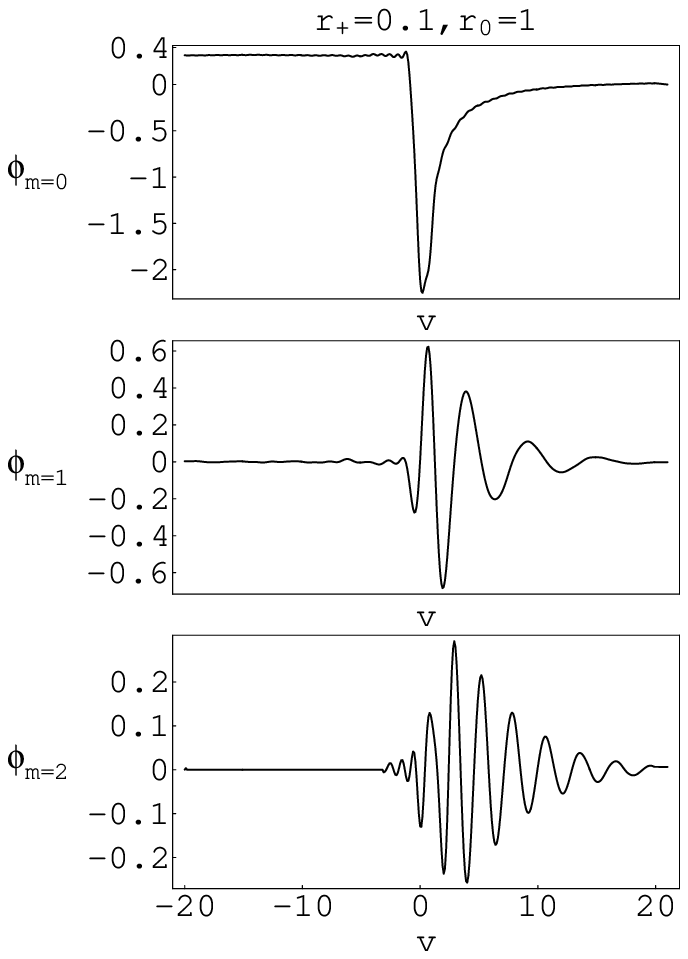}} 
{\noindent {\small Figure 2. Waveforms $\phi(v)$ 
for a  $r_+=0.1\,,\,\,r_0=1$ BTZ black hole, 
for the three
lowest values of $m$. 
}} 
\vskip 3mm
For example, for $m=0$, $\omega_{\rm QNM}=-0.2i(n+1)$, one expects to find
a purely decaying perturbation. This is evident from Fig. 2.  For
$m=1$, $\omega_{\rm QNM}= 1 -0.2i(n+1)$, so the signal should ring (at
late times) with frequency one. This is also clearly seen from Fig 2.
For $m=2$ we have the same kind of behavior. 
For large negative $v$ and fixed $t$ one has large negative $r_*$, 
so one is near the horizon. Thus $\phi(v\rightarrow-\infty)$ 
in Fig. 2 should give the 
same values  as $\phi_0$ at $r_+$ in Fig. 1, which is the case.
The energy spectra peaks at higher $\omega$ when compared to the
fundamental $\omega_{\rm QNM}$ as is evident from Fig. 3, which means
that higher modes are excited.  The total radiated energy as a
function of $m$ goes to zero slower than $1/m$ implying that the total
radiated energy diverges.  However, this divergence can be normalized
by taking a minimum size $L$ for the particle with a cut off given by
$m_{\rm max}\sim \frac{\pi}{2}\frac{r_+}{L}$ \cite{zerillietal}. We
have calculated for $r_+=0.1$ the total energy for the cases
$m=0,1,2$, yielding $E_{m=0}\simeq26$, $E_{m=1}\simeq12$,
$E_{m=2}\simeq 6$. An estimation of the total energy for a
particle with $m_{\rm max}\simeq 1000$ yields $E_{\rm total}\simeq
80$ (the energy is measured in units of $q_0^2m_0^2$).
\vskip 3mm
\centerline{\epsffile{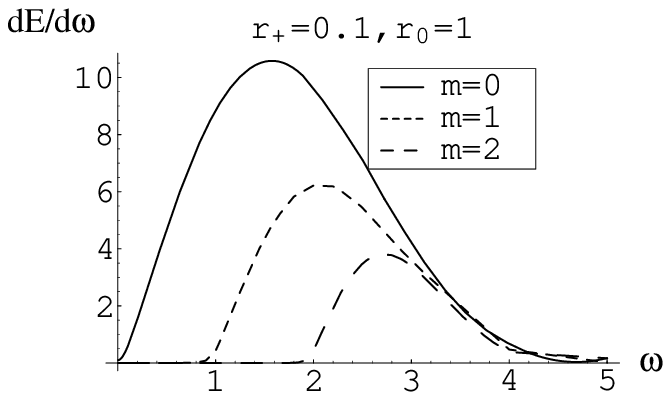}} 
{\noindent {\small Figure 3. Typical energy spectra, here
shown for $r_+=0.1$, and $r_0=1$.
}} 
\vskip 3mm
We have also computed the radiated energy for several values of $r_0$
and verified that is not a monotonic function of $r_0$. For small
values of $r_0$ the energy radiated is a linear function of $(r_0-r_+)$,
for intermediate $r_0$ it has several peaks, and  it grows monotonically 
for large $r_0$.
The zero frequency
limit (ZFL), depends only on the initial data and one can prove that
it is given by $(\frac{dE}{d\omega})_{\omega\rightarrow 0}=
\phi_{0}^2$.  This is to be contrasted to the ZFL for outgoing
gravitational radiation in asymptotically flat spacetimes \cite{smarr}
where it depends only on the initial velocity of the test particle.

%%%%%%%%%%%%%%%%%%%%%%%%%%%%%%%%%%%%%%%%%%%%%%%%%
\noindent
\section{ Conclusions }
\vskip 3mm
%%%%%%%%%%%%%%%%%%%%%%%%%%%%%%%%%%%%%%%%%%%%%%%%%

In conclusion, we have obtained for the first time the plunging waveforms, 
the power spectrum,
and the energy radiated for the collision of a
scalar particle with the BTZ black hole.  We have shown these 
quantities for small black holes. For large black holes the 
results are qualitatively the same, with the one difference that 
the ringing is much shorter (these results together with results 
for higher dimensional black holes will be reported elsewhere).
For the AdS/CFT
correspondence we have added to previous works the precise evolution of
an infalling probe and its radiation. This has implications in the
strongly coupled CFT: to the black hole corresponds a thermal
bath, to the infalling probe corresponds an expanding bubble, and to
the scalar field waves correspond particles decaying into bosons of
the associate operator of the gauge theory.  Both the bubble and the
particles in the CFT thermalize with the characteristic timescale
calculated through the gravity in the bulk 
$1/{\rm Im}[\omega_{\rm QNM}]$, 
oscillate according to Fig. 2, which for late times 
yields the oscillation frequency ${\rm Re}[\omega_{\rm QNM}]$, and
radiate according to Fig. 3.  This is hard to calculate by direct
means in the strongly coupled regime of the gauge theory.

\section*{Acknowledgments} 
We thank Amaro Rica da Silva for discussions and Observat\'orio
Nacional - Rio de Janeiro for hospitality.  This work was partially
funded by FCT-Portugal  through
project PESO/PRO/2000/4014 and through PRAXIS XXI programme.

\end{document}